\documentclass{article}
\usepackage{dcase2021,amsmath,url,times,booktabs,tabularx,amsmath,amssymb}
\usepackage{array,multirow}
\usepackage{arydshln}
\usepackage{multicol}
\usepackage{footnote}
\usepackage{float}
\usepackage{cite}
\usepackage{enumitem}

\usepackage[pdftex]{graphicx}

\newcolumntype{C}[1]{>{\centering\arraybackslash}p{#1}}
\newcolumntype{L}[1]{>{\raggedright\arraybackslash}p{#1}}
\newcolumntype{R}[1]{>{\raggedleft\arraybackslash}p{#1}}

\newcommand{\vect}[1]{{\mbox{\boldmath $#1$}}}

\title{DESCRIPTION AND DISCUSSION ON DCASE 2021 CHALLENGE TASK 2: UNSUPERVISED ANOMALOUS SOUND DETECTION FOR MACHINE CONDITION MONITORING UNDER DOMAIN SHIFTED CONDITIONS}

\name{
Yohei Kawaguchi$^{1}$,
Keisuke Imoto$^{2}$,
Yuma Koizumi$^{3}$,
Noboru Harada$^{4}$,
Daisuke Niizumi$^{4}$,
}
\secondlinename{
Kota Dohi$^{1}$,
Ryo Tanabe$^{1}$,
Harsh Purohit$^{1}$,
and Takashi Endo$^{1}$
}
\address{
$^1$ Hitachi, Ltd., Japan, \url{yohei.kawaguchi.xk@hitachi.com}\\
$^2$ Doshisha University, Japan, \url{keisuke.imoto@ieee.org}\\
$^3$ Google, Japan, \url{koizumiyuma@google.com}\\
$^4$ NTT Corporation, Japan, \url{noboru.harada.pv@hco.ntt.co.jp}\\
}

\begin{document}

\ninept
\maketitle

\begin{sloppy}

\begin{abstract}
We present the task description and discussion on the results of the DCASE 2021 Challenge Task 2. 
In 2020, we organized an unsupervised anomalous sound detection (ASD) task, identifying whether a given sound was normal or anomalous without anomalous training data. 
In 2021, we organized an advanced unsupervised ASD task \textit{under domain-shift conditions}, which focuses on the inevitable problem of the practical use of ASD systems. 
The main challenge of this task is to detect unknown anomalous sounds where the acoustic characteristics of the training and testing samples are different, i.e., domain-shifted. 
This problem frequently occurs due to changes in seasons, manufactured products, and/or environmental noise. 
We received 75 submissions from 26 teams, and several novel approaches have been developed in this challenge. 
On the basis of the analysis of the evaluation results, 
we found that there are two types of remarkable approaches that TOP-5 winning teams adopted: 1) ensemble approaches of ``\textit{outlier exposure}'' (OE)-based detectors and ``\textit{inlier modeling}'' (IM)-based detectors and 2) approaches based on IM-based detection for features learned in a machine-identification task.

\end{abstract}


\section{Introduction}
\label{sec:intro}

Anomalous sound detection (ASD)~\cite{koizumi2017neyman, kawaguchi2017how, koizumi2019neyman, kawaguchi2019anomaly, koizumi2019batch, suefusa2020anomalous, purohit2020deep} is the task of identifying whether the sound emitted from a machine is normal or anomalous. Automatic detection of mechanical failure is an essential technology in the fourth industrial revolution, which includes artificial intelligence (AI)--based factory automation, and also prompt detection of machine anomalies by observing its sounds may be useful for machine condition monitoring.

We organized ``\textit{unsupervised ASD}'' as Task 2 of the Detection and Classification of Acoustic Scenes and Events (DCASE) 2020 Challenge~\cite{koizumi2020dcase} for to connect academic tasks and real-world problems. The main challenge of this task was to detect unknown anomalous sounds under the condition that only normal sound samples have been provided as training data~\cite{koizumi2017neyman, kawaguchi2017how, koizumi2019neyman, kawaguchi2019anomaly, koizumi2019batch, suefusa2020anomalous, purohit2020deep}. In real-world factories, actual anomalous sounds rarely occur but are highly diverse. Therefore, exhaustive patterns of anomalous sounds are impossible to collect. This means that we must detect unknown anomalous sounds that were not in the given training data. 
This unique and real-world oriented task attracted the interest of many participants, 
and resulted in 117 entries from 40 teams, which included several new approaches~\cite{giri2020self, kapka2020id, primus2020anomalous, inoue2020detection}.

For the DCASE 2021 Challenge, we organized a follow-up unsupervised ASD task under domain-shift conditions, which simulates a more challenging issue in real-world applications. The main challenge of this task is that the acoustic characteristics of the training and testing phase are different due to changes in the normal condition such as motor speed and signal-to-noise ratio (SNR). A frequent real-world example of this problem is that the motor speed in a conveyor for transporting products varies in response to product demand; for a product whose demand changes with the seasons, training data recorded in the summer was 300--400 rotations per minute (RPM) (i.e. source domain), but the demand drops in the winter resulting in the motor speed decreasing to 100--200 RPM (i.e, target domain). Because a normal motor sound at 100 RPM is an unknown sound for the ASD system, it could incorrectly be detected as an anomalous sound. Therefore, methods to deal with such drift in normal conditions are required to accelerate the real-world application of ASD.

As the first benchmark task for domain-shift problems in ASD, we designed the DCASE Challenge 2021 Task 2 
``\textit{Unsupervised Detection of Anomalous Sounds for Machine Condition Monitoring under Domain-Shifted Conditions}.'' 
The scope includes differences in operating speed, machine load, environmental noise, and so on. 
After briefly introducing this task, we discuss remarkable approaches and their potential problems on the basis of the analysis of all 75 submissions from 26 teams.

\section{Unsupervised Anomalous Sound Detection under Domain-Shifted Conditions} 
\label{sec:uasd}

Let the $L$-sample time-domain observation $\vect{x} \in \mathbb{R}^L$ be an audio clip that includes a sound emitted from a machine. ASD is the determination of whether a machine is in a normal or anomalous state from $\vect{x}$. To determine the state of the machine, an anomaly score is calculated; it takes a large value when the machine is anomalous, and vice versa. To calculate the anomaly score, we have to prepare an anomaly score calculator $\mathcal{A}$ with parameter $\theta$. The input of $\mathcal{A}$ is the audio clip $\vect{x}$ and additional information given by its file path, and one anomaly score $\mathcal{A}_{\theta}(\vect{x})$ is output. Then, the machine is determined to be anomalous when the anomaly score $\mathcal{A}_{\theta}(\vect{x})$ exceeds a pre-defined threshold value $\phi$ as
\begin{equation}
\mbox{Decision} = \left\{
\begin{array}{ll}
\mbox{Anomaly} & (\mathcal{A}_{\theta}(\vect{x}) > \phi)\\
\mbox{Normal} & (\mbox{otherwise}).
\end{array}
\right.
\label{eq:det}
\end{equation}
The primal difficulty in this task is to train $\mathcal{A}$ so that $\mathcal{A}_{\theta}(\vect{x})$ becomes a large value when the machine is anomalous, even though only normal sounds are available as training data.

In addition to the regular unsupervised ASD, we have to solve the domain-shift problem in real-world cases. As mentioned in Section \ref{sec:intro}, domain shifts refer to the difference in conditions between training and testing phases. The conditions differ in operating speed, machine load, viscosity, heating temperature, environmental noise, SNR, etc. The difference in conditions causes a gap in the sound characteristics, i.e., the distribution of the observation in the feature space changes. Here, two conditions are defined: \textbf{source domain} and \textbf{target domain}. The source domain refers to the original condition with a sufficient number of training clips, and the target domain refers to another state that has changed from the source domain. Let $\mathcal{D}_S$, $\mathcal{D}_T$, and $\mathcal{D}_{TA}$ be the distributions of $\vect{x}$ under the normal condition in the source domain, the normal condition in the target domain, and the anomalous condition in the target domain, respectively. 
The regular unsupervised ASD task is to determine whether $\vect{x}$ was generated from $\mathcal{D}_T$ or $\mathcal{D}_{TA}$ under the condition that clips from $\mathcal{D}_{T} (= \mathcal{D}_{S})$ are available as training data, but clips from $\mathcal{D}_{TA}$ are not. On the other hand, in the domain shift scenario, detection must be performed under the condition that clips from $\mathcal{D}_{T} (\ne \mathcal{D}_{S})$ are available as training data, but clips from $\mathcal{D}_{TA}$ are not. Note that only a few clips from $\mathcal{D}_{T}$ are provided as training data in 2021's task setting.

\section{Task Setup} 
\label{sec:task}

\subsection{Dataset} 
\label{sec:dataset}

The data used for this task comprises parts of the ToyADMOS2~\cite{harada2021toyadmos2} and MIMII DUE~\cite{tanabe2021mimiidue} datasets
consisting of the normal/anomalous operating sounds of seven types of toy/real machines.
We intentionally damaged machines to collect the anomalous sounds in these datasets.
We provide the following types of machines: 
ToyCar and ToyTrain from ToyADMOS2, and fan, gearbox, pump, slide rail, and valve from MIMII DUE.
To simplify the task, we use only the first channel of multichannel recordings; all recordings can be regarded as the single-channel recordings of a fixed microphone.
Each recording is 10-sec-long audio that includes both the machine's operating sound and environmental noise. 
The sampling rate of all signals is 16 kHz. 
We mixed machine sounds with environmental noise, and only noisy recordings are available as training/test data. 
The environmental noise samples were recorded in several real factory environments. 
For the details of the recording procedure, please refer to the papers on ToyADMOS2~\cite{harada2021toyadmos2} and MIMII DUE~\cite{tanabe2021mimiidue}.

In this task, we define two important terms: \textbf{machine type} and \textbf{section}.
\begin{description}[style=unboxed,leftmargin=0cm]
\item [Machine type] refers to the type of machine, which can be one of seven in this task: fan, gearbox, pump, slide rail, ToyCar, ToyTrain, and valve.
\item [Section] is defined as a subset of the data within one machine type and consists of data from the source and target domains. 
A section is a unit for calculating performance metrics and is almost identical to ``machine ID'' in the 2020 version. 
In the 2020 version, there was a one-to-one correspondence between machine IDs and products, 
but in the 2021 version, machines of the same product appear in different sections (Sections 00--02 of the gearbox are the same product, and sections 03--04 of the gearbox are the same product.), and multiple products appear in the same section (Section 01 of the fan contains two products \cite{tanabe2021mimiidue}).
\end{description}

We provide three datasets: 
\textbf{development dataset}, \textbf{additional training dataset}, and \textbf{evaluation dataset}. 
\begin{description}[style=unboxed,leftmargin=0cm]
\item [Development dataset] consists of three sections for each machine type (Sections 00, 01, and 02), and each section is a complete set of training and test data. 
For each section, this dataset provides 
(i) around 1,000 clips of normal sounds in a source domain for training, 
(ii) only three clips of normal sounds in a target domain for training, 
(iii) around 100 clips of both normal and anomalous sounds in the source domain for the test, and 
(iv) around 100 clips each of normal and anomalous sounds in the target domain for the test.
\item [Additional training dataset] provides the other three sections for each machine type (Sections 03, 04, and 05). 
Each section consists of 
(i) around 1,000 clips of normal sounds in a source domain for training and 
(ii) only three clips of normal sounds in a target domain for training. 
\item [Evaluation dataset] provides test clips for the three sections (Sections 03, 04, and 05) identical to those in the additional training dataset. 
Each section consists of 
(i) test clips in the source domain and 
(ii) test clips in the target domain, none of which have a condition label (i.e., normal or anomaly). 
Note that the sections of the evaluation dataset (Sections 03, 04, and 05) are different from the development dataset (Sections 00, 01, and 02).
\end{description}

\subsection{Evaluation metrics} 
\label{sec:metrics}

The area under the curve (AUC) and partial-AUC (pAUC) for receiver operating characteristic (ROC) curves are used for evaluation as well as the 2020 edition \cite{koizumi2020dcase}. 
The pAUC is an AUC calculated from a portion of the ROC curve over the pre-specified range of interest. 
In our metric, the pAUC is calculated as the AUC over a low false-positive-rate (FPR) range $\left[ 0, p \right]$. 
The AUC and pAUC for each machine type, section, and domain are defined as
\begin{align}
{\rm AUC}_{m, n, d} &= \frac{1}{N_{-}N_{+}} \sum_{i=1}^{N_{-}} \sum_{j=1}^{N_{+}} \mathcal{H} (\mathcal{B}_{\theta, j, i}),\\
{\rm pAUC}_{m, n, d} &= \frac{1}{\lfloor p N_{-} \rfloor N_{+}} \sum_{i=1}^{\lfloor p N_{-} \rfloor} \sum_{j=1}^{N_{+}} \mathcal{H} (\mathcal{B}_{\theta, j, i}),
\end{align}
where 
$\mathcal{B}_{\theta, j, i} = \mathcal{A}_{\theta} (x_{j}^{+}) - \mathcal{A}_{\theta} (x_{i}^{-})$, $m$ represents the index of a machine type,
$n$ represents the index of a section,
$d = \{ {\rm source}, {\rm target} \}$ represents a domain,
$\lfloor \cdot \rfloor$ is the flooring function,
and $\mathcal{H} (x)$ returns 1 when $x > 0$ and 0 otherwise.
$\{x_{i}^{-}\}_{i=1}^{N_{-}}$ and $\{x_{j}^{+}\}_{j=1}^{N_{+}}$ are normal and anomalous test clips in domain $d$ in section $n$ in machine type $m$, respectively,
and they have been sorted so that their anomaly scores are in descending order.
$N_{-}$ and $N_{+}$ are the number of normal and anomalous test clips in domain $d$ in section $n$ in machine type $m$, respectively.
The additional use of the pAUC is based on practical requirements.
If an ASD system frequently gives false alarms, we cannot trust it.
Therefore, it is important to increase the true-positive rate under low FPR conditions. In this task, we will use $p=0.1$.
The official score $\Omega$ for each submitted system is given by the harmonic mean of the AUC and pAUC scores over all machine types, all sections, and both domains.
As the aforementioned equations show, a threshold value does not need to be determined to calculate AUC, pAUC, or the official score because the threshold value is set to the anomaly score of a normal test clip.

\subsection{Baseline systems and results}
\label{sec:baseline}

The task organizers provided two baseline systems. 

\vspace{6pt}
\noindent
\textbf{Autoencoder-based baseline:}
The first baseline is an autoencoder (AE)-based anomaly score calculator and the same as the DCASE 2020 task 2. 
Details are described in the 2020 task description~\cite{koizumi2020dcase}.
The anomaly score $\mathcal{A}_{\theta}$ is calculated as the mean square error of reconstruction for the observed sound.
To obtain small anomaly scores for normal sounds, the AE is trained to minimize the reconstruction error of the normal training data. 
This method is based on the assumption that the AE cannot reconstruct sounds that are not used in training, that is, unknown anomalous sounds.
If and only if $A_{\theta}$ for each test clip is greater than a threshold, the clip is judged to be anomalous.

\vspace{6pt}
\noindent
\textbf{``\textit{Outlier exposure}''-based baseline using MobileNetV2:}
The second baseline is an anomaly score calculator obtained by using an approach called ``\textit{outlier exposure}'' (OE)~\cite{hendrycks2019deep} for the machine-identification task.
In DCASE 2020, 10 of the 40 teams used this approach~\cite{giri2020self, primus2020anomalous, inoue2020detection, Zhou2020, Lopez2020, Wilkinghoff2020, Shinmura2020, Wei2020, Ahmed2020, Xiao2020}, and with four of them ~\cite{giri2020self, Zhou2020, Shinmura2020, Wei2020} using the model of MobileNetV2~\cite{sandler2018mobilenetv2}.
The models of this baseline are trained to identify from which section the observed signal was generated; it outputs the softmax value that is the predicted probability for each section. 
The anomaly score is calculated as the averaged negative logit of the predicted probabilities for the correct section.
We first calculate the log-mel-spectrogram of the input $X = \{X_t\}_{t = 1}^T$, 
where $X_t \in \mathbb{R}^F$, and $F$ and $T$ are the number of mel-filters and time-frames, respectively.
Then, the acoustic feature (two-dimensional image) at $t$ is obtained by concatenating consecutive frames of the log-mel-spectrogram as $\psi_t = (X_t, \cdots, X_{t + P - 1}) \in \mathbb{R}^{P \times F}$.
By shifting the context window by $L$ frames, $B (= \lfloor \frac{T - P}{L} \rfloor)$ images are extracted.
The frame size of the short-time Fourier transform (STFT) is 64 ms, and the hop size is 50 \%. 
In addition, $F=128$, $P=64$, and $L=8$.
The Adam optimizer is used, and we fix the learning rate to 0.00001. 
We stop the training process after 20 epochs, and the batch size is 32. 
We train models independently for each machine type using normal clips from all sections of that machine type.
The sections are used as classes to train the individual models.
The anomaly score is calculated as:
\begin{equation}
\mathcal{A}_{\theta}(X) = \frac{1}{B} \sum_{b = 1}^B \log \left( \frac{1 - p_{\theta}(\psi_{t(b)})}{p_{\theta}(\psi_{t(b)})} \right),
\end{equation}
where $t(b)$ is the beginning frame index of the $b$-th image and $p_{\theta}$ is the softmax output by MobileNetV2 for the correct section.

\begin{figure*}[t!]
\begin{center}
\includegraphics[width=1.0\hsize,clip]{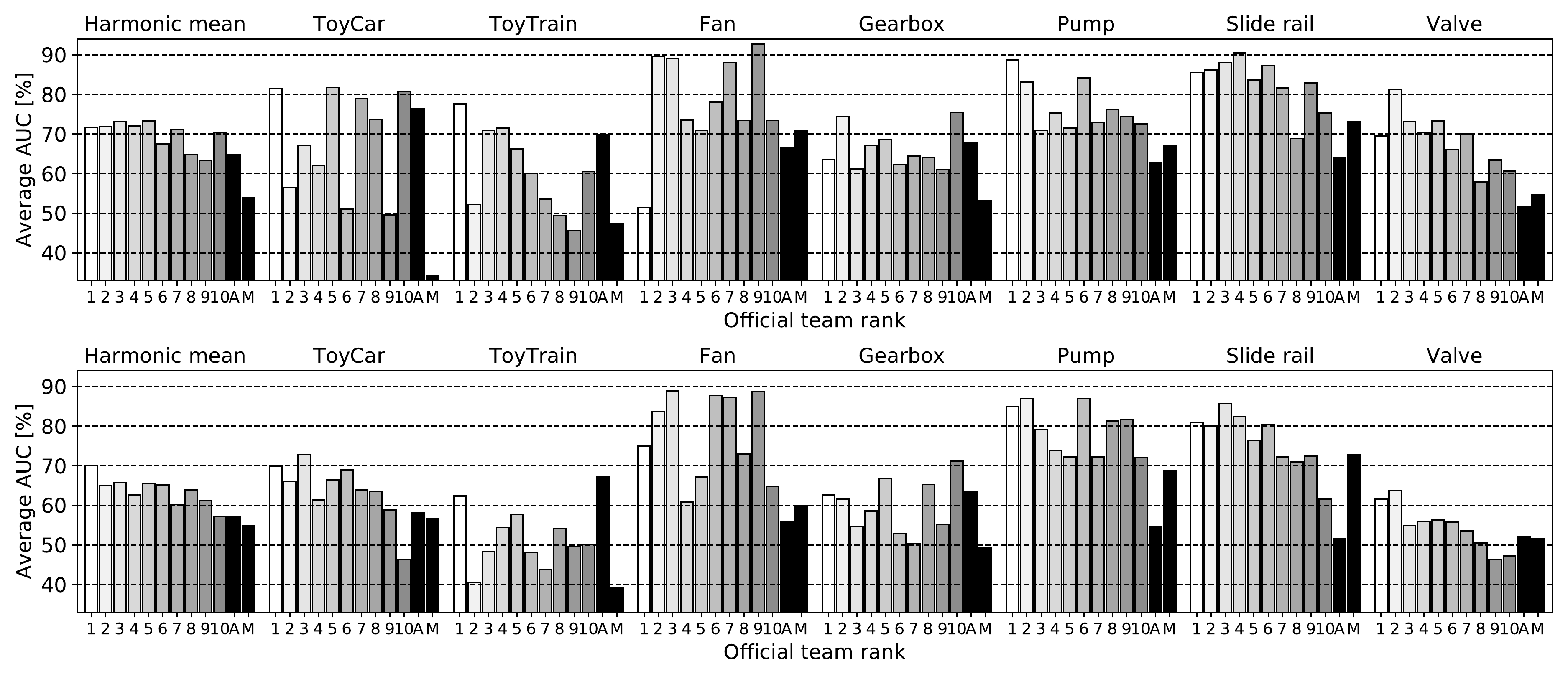}\\
\caption{Evaluation results of top 10 teams in team ranking. 
AUC for source domains (top) and AUC for target domains (bottom). 
Label ``A'' and ``M'' on the x-axis means AE-based and MobileNetV2-based baselines, respectively.}
\label{fig:auc}
\end{center}
\end{figure*}


\setlength{\tabcolsep}{1mm} 
\begin{table}[t]
\begin{center}
\caption{Results of the AE-based baseline}
\label{tab:ae_results}
\scriptsize
\begin{tabular}{l l c c p{1pt} c c}
\hline
\multicolumn{2}{@{}l}{\multirow{2}{*}{ \begin{tabular}{@{\hskip0pt}l@{\hskip0pt}} Section \end{tabular} }} &
\multicolumn{2}{c}{AUC [\%]} &&
\multicolumn{2}{c}{pAUC [\%]} \\
\cline{3-4} \cline{6-7}
& &
\multicolumn{1}{c}{Source} &
\multicolumn{1}{c}{Target} &&
\multicolumn{1}{c}{Source} &
\multicolumn{1}{c}{Target} \\
\hline
  \multirow{3}{*}{ToyCar}
  &	00 & $67.63 \pm 1.21$ & $54.50 \pm 0.89$ && $51.87 \pm 0.50$ & $50.52 \pm 0.20$ \\
  & 	01 & $61.97 \pm 1.50$ & $64.12 \pm 1.07$ && $51.82 \pm 0.87$ & $52.14 \pm 0.80$ \\
  & 	02 & $74.36 \pm 0.82$ & $56.57 \pm 1.53$ && $55.56 \pm 0.83$ & $52.61 \pm 1.20$ \\ 
  \hline

  \multirow{3}{*}{ToyTrain} 
  &	00 & $72.67 \pm 1.19$ & $56.07 \pm 0.80$ && $69.38 \pm 1.06$ & $50.62 \pm 0.68$ \\
  & 	01 & $72.65 \pm 0.32$ & $51.13 \pm 0.53$ && $62.52 \pm 0.88$ & $48.60 \pm 0.13$ \\
  & 	02 & $69.91 \pm 0.33$ & $55.57 \pm 1.07$ && $47.48 \pm 0.02$ & $50.79 \pm 0.93$ \\ 
  \hline

  \multirow{3}{*}{Fan} 
  &	00 & $66.69 \pm 0.81$ & $69.70 \pm 0.32$ && $57.08 \pm 0.15$ & $55.13 \pm 0.34$ \\
  & 	01 & $67.43 \pm 1.12$ & $49.99 \pm 0.48$ && $50.72 \pm 0.42$ & $48.49 \pm 0.38$ \\
  & 	02 & $64.21 \pm 1.27$ & $66.19 \pm 1.23$ && $53.12 \pm 0.78$ & $56.93 \pm 1.37$ \\ 
  \hline

  \multirow{3}{*}{Gearbox} 
  &	00 & $56.03 \pm 0.53$ & $74.29 \pm 0.51$ && $51.59 \pm 0.16$ & $55.67 \pm 0.97$ \\
  & 	01 & $72.77 \pm 0.72$ & $72.12 \pm 1.06$ && $52.30 \pm 0.18$ & $51.78 \pm 0.15$ \\
  & 	02 & $58.96 \pm 0.53$ & $66.41 \pm 0.72$ && $51.82 \pm 0.29$ & $53.66 \pm 0.57$ \\ 
  \hline

  \multirow{3}{*}{Pump} 
  &	00 & $67.48 \pm 0.58$ & $58.01 \pm 0.57$ && $61.83 \pm 0.41$ & $51.53 \pm 0.27$ \\
  & 	01 & $82.38 \pm 0.27$ & $47.35 \pm 0.53$ && $58.29 \pm 0.77$ & $49.65 \pm 1.46$ \\
  & 	02 & $63.93 \pm 0.45$ & $62.78 \pm 0.70$ && $55.44 \pm 0.52$ & $51.67 \pm 0.35$ \\ 
  \hline

  \multirow{3}{*}{Slide rail} 
  &	00 & $74.09 \pm 0.48$ & $67.22 \pm 0.45$ && $52.45 \pm 0.63$ & $57.32 \pm 0.52$ \\
  & 	01 & $82.16 \pm 0.35$ & $66.94 \pm 0.39$ && $60.29 \pm 0.30$ & $53.08 \pm 0.39$ \\
  & 	02 & $78.34 \pm 0.16$ & $46.20 \pm 0.77$ && $65.16 \pm 0.55$ & $50.10 \pm 0.31$ \\ 
  \hline

  \multirow{3}{*}{Valve} 
  &	00 & $50.34 \pm 0.27$ & $47.12 \pm 0.18$ && $50.82 \pm 0.16$ & $48.68 \pm 0.09$ \\
  & 	01 & $53.52 \pm 0.33$ & $56.39 \pm 1.42$ && $49.33 \pm 0.10$ & $53.88 \pm 0.61$ \\
  & 	02 & $59.91 \pm 0.34$ & $55.16 \pm 0.22$ && $51.96 \pm 0.52$ & $48.97 \pm 0.04$ \\ 
  \hline

\end{tabular}
\end{center}
\end{table}

\setlength{\tabcolsep}{1mm} 
\begin{table}[t]
\begin{center}
\caption{Results of the MobileNetV2-based baseline}
\label{tab:mob_results}
\scriptsize
\begin{tabular}{l l c c p{1pt} c c}
\hline
\multicolumn{2}{@{}l}{\multirow{2}{*}{ \begin{tabular}{@{\hskip0pt}l@{\hskip0pt}} Section \end{tabular} }} &
\multicolumn{2}{c}{AUC [\%]} &&
\multicolumn{2}{c}{pAUC [\%]} \\
\cline{3-4} \cline{6-7}
& &
\multicolumn{1}{c}{Source} &
\multicolumn{1}{c}{Target} &&
\multicolumn{1}{c}{Source} &
\multicolumn{1}{c}{Target} \\
\hline
  \multirow{3}{*}{ToyCar}
  &	00 & $66.56 \pm 2.68$ & $61.32 \pm 5.94$ && $66.47 \pm 5.67$ & $52.61 \pm 2.41$ \\
  & 	01 & $71.58 \pm 5.54$ & $72.48 \pm 3.68$ && $66.44 \pm 2.84$ & $63.99 \pm 2.60$ \\
  & 	02 & $40.37 \pm 7.19$ & $45.17 \pm 3.36$ && $47.48 \pm 0.23$ & $48.85 \pm 0.94$ \\ 
  \hline

  \multirow{3}{*}{ToyTrain} 
  &	00 & $69.84 \pm 4.39$ & $46.28 \pm 3.85$ && $54.43 \pm 1.65$ & $51.27 \pm 0.73$ \\
  & 	01 & $64.79 \pm 3.65$ & $53.38 \pm 2.47$ && $54.09 \pm 1.15$ & $49.60 \pm 0.88$ \\
  & 	02 & $69.28 \pm 6.73$ & $51.42 \pm 2.64$ && $47.66 \pm 0.40$ & $53.40 \pm 1.12$ \\ 
  \hline

  \multirow{3}{*}{Fan} 
  &	00 & $43.62 \pm 2.35$ & $53.34 \pm 2.03$ && $50.45 \pm 1.15$ & $56.01 \pm 1.38$ \\
  & 	01 & $78.33 \pm 1.52$ & $78.12 \pm 4.25$ && $78.37 \pm 2.26$ & $66.41 \pm 7.16$ \\
  & 	02 & $74.21 \pm 3.85$ & $60.35 \pm 3.79$ && $76.80 \pm 0.78$ & $60.97 \pm 6.55$ \\ 
  \hline

  \multirow{3}{*}{Gearbox} 
  &	00 & $81.35 \pm 1.59$ & $75.02 \pm 2.92$ && $70.46 \pm 3.67$ & $64.77 \pm 2.52$ \\
  & 	01 & $60.74 \pm 5.11$ & $56.27 \pm 8.27$ && $53.88 \pm 2.82$ & $53.30 \pm 2.97$ \\
  & 	02 & $71.58 \pm 7.16$ & $64.45 \pm 9.67$ && $62.23 \pm 6.67$ & $55.58 \pm 7.90$ \\
  \hline

  \multirow{3}{*}{Pump} 
  &	00 & $64.09 \pm 4.34$ & $59.09 \pm 3.08$ && $62.40 \pm 1.90$ & $53.96 \pm 0.93$ \\
  & 	01 & $86.27 \pm 3.18$ & $71.86 \pm 5.97$ && $66.66 \pm 5.23$ & $62.69 \pm 2.33$ \\
  & 	02 & $53.70 \pm 4.99$ & $50.16 \pm 3.78$ && $50.98 \pm 1.23$ & $51.69 \pm 1.03$ \\ 
  \hline

  \multirow{3}{*}{Slide rail} 
  &	00 & $61.51 \pm 4.92$ & $51.96 \pm 3.17$ && $53.97 \pm 2.03$ & $51.96 \pm 2.96$ \\
  & 	01 & $79.97 \pm 3.70$ & $46.83 \pm 10.65$ && $55.62 \pm 1.57$ & $52.02 \pm 4.17$ \\
  & 	02 & $79.86 \pm 1.41$ & $55.61 \pm 5.48$ && $71.88 \pm 4.64$ & $55.71 \pm 2.84$ \\
  \hline

  \multirow{3}{*}{Valve} 
  &	00 & $58.34 \pm 4.01$ & $52.19 \pm 3.33$ && $54.97 \pm 4.43$ & $51.54 \pm 1.88$ \\
  & 	01 & $53.57 \pm 2.26$ & $68.59 \pm 2.84$ && $50.09 \pm 0.45$ & $57.83 \pm 2.49$ \\
  & 	02 & $56.13 \pm 1.96$ & $53.58 \pm 0.55$ && $51.69 \pm 0.32$ & $50.86 \pm 0.84$ \\
  \hline

\end{tabular}
\end{center}
\end{table}

Tables \ref{tab:ae_results} and \ref{tab:mob_results} show the AUC and pAUC scores for the two baselines, respectively. 
Because the results produced with a GPU are generally non-deterministic, the average and standard deviations from these five independent trials (training and testing) are shown.

\section{Challenge Results and Discussion}

\subsection{Results for evaluation dataset}

We received 75 submissions from 26 teams, and 20 teams achieved better performance than the baseline systems. 
The harmonic means of the AUC scores of the top 10 teams~\cite{LopezIL2021, MoritaSECOM2021, WilkinghoffFKIE2021, KuroyanagiNU-HDL2021, SakamotoFixstars2021, ZhouPSH2021, HeXJU2021, TozickaNSW2021, CaiSMALLRICE2021, NaritaAIT2021} 
are shown in Fig. \ref{fig:auc} for the source and target domains. 
As shown in the figure, the performance for which machine type is high or low varies greatly from team to team. 
However, the score on the target domain roughly correlates to the official ranking.

We find that there are two remarkable approaches in high-rank solutions: 
the first is an ensemble of OE-based detection and ``\textit{inlier modeling}'' (IM)-based detection~\cite{LopezIL2021, KuroyanagiNU-HDL2021, SakamotoFixstars2021}.
Here, IM refers to out-of-distribution (OOD) detection methods based on modeling a distribution of inlier samples, for example, 
AE, k-nearest neighbors (kNN), local outlier factor (LOF), Gaussian mixture models (GMM), normalizing flows (NF), interpolation deep neural network (IDNN)~\cite{suefusa2020anomalous}, and their conditional versions.
The second approach is IM-based detection for features learned in a machine-identification task~\cite{MoritaSECOM2021, WilkinghoffFKIE2021}.
We describe the details in the following sections.

\subsection{Parallel-type hybrid approach: ensemble of OE-based and IM-based detectors}

The first, fourth, and fifth-place teams~\cite{LopezIL2021, KuroyanagiNU-HDL2021, SakamotoFixstars2021} utilized this type of approach.
OE-based detectors have a weakness in that their performance is severely degraded when the distributions of different sections (roughly, machine products) are too similar or too different~\cite{koizumi2020dcase, dohi2021flow}; 
ensembles of OE and IM can reduce this OE weakness by leveraging the robustness of IM.
The 2020 first-place team~\cite{giri2020self} used an ensemble of OE-based and IM-based detectors, using MobileNetV2 for OE and IDNN for IM. 
The 2021 first-place team~\cite{LopezIL2021} also used this type of approach, using multiple types of OE models and a conditional NF. 
Surprisingly, the technical report shows that this team did not perform any domain adaptation, but the performance of the target domain is outstanding. 
Taking into account the low performance of the individual subsystems of this team, we can guess that the ensemble gave them high generalization performance in both domains.

The fourth and fifth-place teams~\cite{KuroyanagiNU-HDL2021, SakamotoFixstars2021} also took ensembles of OE and IM. 
However, unlike the first-place team, they prepared a model for each domain and performed domain adaptation, resulting in comparable AUC scores to the second and third-place teams in both domains.
For example, the fifth-ranked team~\cite{SakamotoFixstars2021} is the only team to achieve AUC scores over 55\% on all machine types in both domains.
Although this ensemble-type approach tends to increase its model complexity, 
surprisingly, the model of the fifth-place team~\cite{SakamotoFixstars2021} is small.
Further improving the performance of domain adaptation while maintaining the compactness of the model will continue to be a research problem.



\subsection{Serial-type hybrid approach: IM-based detection for features learned in a machine-identification task}

The second and third-place teams~\cite{MoritaSECOM2021, WilkinghoffFKIE2021} utilized this type of approach.
They extracted features for the machine-identification task and performed IM-based detection on the extracted features. 
In training, the feature extraction model was first trained in the machine-identification task like OE-based methods, and then the IM-based detection model was trained.
The aforementioned ensemble-type approach can be thought of as a parallel-type hybrid, whereas this approach can be thought of as a serial-type hybrid of OE and IM.
This approach uses the powerful feature extraction of OE, but overcomes its aforementioned instability by taking advantage of the robustness of IM.
In addition, this approach has the advantage of preventing the model complexity associated with ensembles.

Domain adaptation was performed only on IM-based detectors and not on feature extractors by the second and third-place teams~\cite{MoritaSECOM2021, WilkinghoffFKIE2021}.
Such a domain adaptation method is less prone to overfitting because it fine-tunes only a limited range, and is considered effective when the number of training samples in the target domain is small.
However, there is no guarantee that the features that are effective for machine identification will remain effective after the domain shift.
In the future, it is desirable to verify how wide the effective range of this approach is and how far its performance for domain adaptation  can be improved.

\section{Conclusion}

We presented an overview of the task and analysis of the solutions submitted to the DCASE 2021 Challenge Task 2. 
The main challenge of this task was to detect unknown anomalous sounds where the acoustic characteristics of the training and testing samples were different. 
We analyzed all evaluation results and submissions, 
and found that there are two types of remarkable approaches that TOP-5 winning teams adopted, i.e., 1) the parallel-type hybrid: ensemble approaches of OE-based and IM-based detectors and 2) the serial-type hybrid: approaches based on IM-based detection for features learned in the machine-identification task. 
Both two approaches are promising, but there is some room for performance improvement for domain adaptation. For the parallel-type hybrid, future work is to improve the performance of domain adaptation while maintaining the compactness of the model. Future work is needed for the serial-type hybrid to verify how wide this approach's effective range is and improve the domain adaptation performance.



\newpage

{\footnotesize
\bibliographystyle{IEEEtran}
\bibliography{refs}
}
\end{sloppy}
\end{document}